\documentstyle[prl,aps,amssymb,twocolumn,epsf]{revtex}

\title{Separability properties of tripartite states with
       $\rm U\!\otimes\! U\!\otimes\! U$-symmetry}
\author{T. Eggeling
\thanks{Electronic Mail:
\tt{T.Eggeling@tu-bs.de}}{{}\quad and\ }  R.~F. Werner
\thanks{Electronic Mail:
\tt{R.Werner@tu-bs.de}}
  \\[1ex]
  {\small Institut f{\"u}r Mathematische Physik, TU Braunschweig,}\\
  {\small Mendelssohnstr.3, 38106 Braunschweig, Germany.}}
\date{March 2, 2000}

\def\Hh{{\cal H}}

\def\Cx{{\Bbb C}}
\def\Rl{{\Bbb R}}
\def\idty{{\openone}}

\def\tr{\mathop{\rm tr}\nolimits}

\def\V#1 {V_{(#1)}} 
\def\pr{{\bf P}}    
\def\wss{{\cal W}} 
\def\wssp{\wss^0}  
\def\tsp{{\cal T}} 
\def\bsp{{\cal B}} 
\def\ppt{{\cal P}}
\def\ket#1{\mid#1\rangle}

\def\ket #1{\vert #1\rangle}
\def\braket #1#2{\langle #1 \vert #2\rangle}
\def\ketbra #1#2{\vert #1\rangle \! \langle #2\vert}

\def\norm #1{\Vert #1\Vert}
\def\ko #1#2{\left[#1,#2\right]_{-}}
\def\Null{{\bf 0}}

\def\pps{{\cal P}}
\def\ptr{^{T_1}}

\newtheorem{Cri}{Criterion}

\begin{document}
\draft
\maketitle

\begin{abstract} We study separability properties in a
$5$-dimensional set of states of quantum systems composed of three
subsystems of equal but arbitrary finite Hilbert space dimension.
These are the states, which can be written as linear combinations
of permutation operators, or, equivalently, commute with unitaries
of the form $U\otimes U\otimes U$. We compute explicitly the
following subsets: (1) triseparable states, which are convex
combinations of triple tensor products, (2) biseparable states,
which are separable for a twofold partition of the system, and (3)
states with positive partial transpose with respect to such a
partition.
\end{abstract}


\narrowtext

\section{Introduction}
One of the difficulties in the theory of entanglement is that
state spaces are  usually fairly high dimensional convex sets.
Therefore, to explore in detail the potential of entangled states
one often has to rely on lower dimensional ``laboratories''. An
example of this was the role played by a one-dimensional family of
bipartite states \cite{W89}, which has come to be known as
``Werner states''. In this paper we present a similar laboratory,
designed for the study of entanglement between three subsystems.
The basic idea is rather similar to \cite{W89}, and we believe
this set shares many of the virtues with its bipartite
counterpart. Firstly, the states have an explicit parametrization
as linear combinations of permutation operators. This is helpful
for explicit computations. Secondly, there is a ``twirl''
operation which brings an arbitrary tripartite state to this
special subset. This proved to be very helpful for the discussion
of entanglement distillation of bipartite entanglement: the first
useful distillation procedures worked by starting with Werner
states, applying a suitable distillation operation, and then the
twirl projection to come back to the simple and well understood
subset, thus allowing iteration. Geometrically this means that the
subset we investigate is both a section of the state space by a
hyperplane and the image of the state space under a projection.
The basic technique for getting such subsets is averaging over a
symmetry group of the entire state space. Such an averaging
projection preserves entanglement if it is an average only over
local (factorizing) unitaries (see \cite{DCT99} for a recent
example different from ours).

The third useful property of the states we study is that they can
be defined for systems of arbitrary finite Hilbert space dimension
$d$, which is again important for the discussion of distillation.
Surprisingly, it even turns out that in the parametrization we
choose all the sets we investigate are also independent of
dimension.

We now describe the entanglement (or separability) properties we
will chart for these special states. Of course, we can split the
system into just two subsystems and apply the usual
separability/entanglement distinctions.  A split $1\vert23$ then
corresponds to the grouping of the Hilbert space
$\Hh_1\otimes\Hh_2\otimes\Hh_3$ into
$\Hh_1\otimes(\Hh_2\otimes\Hh_3)$. We call a density operator
$\rho$ on this Hilbert space {\it$1|23$-separable}
($\rho\in\bsp_1$), or just {\it biseparable} if the partition is
clear from the context, if we can write
\begin{equation}\label{def:bisep}
  \rho=\sum_\alpha \lambda_\alpha\ \rho^{(1)}_\alpha\otimes
          \rho^{(23)}_\alpha,
\end{equation}
with $\lambda_\alpha\geq0$ and density operators $\rho^{(23)}_\alpha$ on
$\Hh_2\otimes\Hh_3.$ Furthermore, as it is a necessary condition
for biseparability (cf. Peres~\cite{Peres}), we are going to look
at those states having a positive partial transpose with regard to
such a split. Recall that the partial transpose $A\mapsto A\ptr$
of operators on $\Hh_1\otimes\Hh_2$ is defined by
\begin{equation}\label{eq:ppt}
  (\sum_\alpha A_\alpha\otimes B_\alpha)\ptr
      = \sum_\alpha A_\alpha^T\otimes B_\alpha,
\end{equation}
where $A^T$ on the right hand side is the ordinary transposition
of matrices with respect to a fixed basis.

As a genuinely ``tripartite'' notion of separability, we consider
states, called {\it triseparable} (or ``three-way classically
correlated''), which can be decomposed as
\begin{equation}\label{def:trisep}
  \rho=\sum_\alpha \lambda_\alpha\ \rho^{(1)}_\alpha\otimes
          \rho^{(2)}_\alpha\otimes\rho^{(3)}_\alpha,
\end{equation}
 where $\lambda_\alpha\geq0$, and the $\rho^{(i)}_\alpha$ are
density operators on the respective Hilbert spaces.

The detailed computations leading to the results presented here
will be published elsewhere, together with the discussion of
further aspects, such as violations of Bell inequalities, and some
distillability relations.

\section{Description of the states}

We consider only the case, where $\Hh_1=\Hh_2=\Hh_3=\Hh$ is
$d$-dimensional ($2\leq d<\infty$). Then the set of states we are
going to study, which will be denoted  by $\wss$, is the set of
density operators, which commute with all unitaries of the form
$U\otimes U\otimes U$. The subsets of $\wss$ we will compute will
be denoted by $\bsp_1$, for the $1\vert23$-biseparable states,
$\ppt_1$ for the states with positive partial transpose with
respect to $1\vert23$, and by $\tsp$ for the triseparable states.
Of course, $\tsp\subset\bsp_1\subset\ppt_1\subset\wss$. We will
see that all these inclusions are strict, even if we take
biseparability with respect to all three partitions:
$\tsp\neq(\bsp_1\cap\bsp_2\cap\bsp_3)$.

It is a fundamental fact from the representation theory of
classical groups \cite{Weyl} that these states are precisely those
which can be written as a linear combination of permutation
operators
\begin{equation}\label{sumpiV}
\rho\in\wss\Leftrightarrow\rho=\sum_\pi\mu_\pi V_\pi
\end{equation}
with coefficients $\mu_\pi\in\Cx$ and the unitary permutation
operators $V_\pi$ defined by
$$ V_\pi\ \phi_1\otimes\phi_2\otimes\phi_3
=\phi_{\pi^{-1}1}\otimes\phi_{\pi^{-1}2}\otimes\phi_{\pi^{-1}3}$$
implementing the permutation symmetry of the three sites. For
density operators hermiticity and normalization reduce the $3!=6$
complex parameters to five real ones (an explicit choice will be
made below).

As usual, the integral
\begin{equation}\label{def:pr}
  \pr \rho
    =\int\!dU\ (U\otimes U\otimes U)\rho\,(U\otimes U\otimes U)^*,
\end{equation}
with respect to the normalized invariant measure ``$dU$'' of the
unitary group ${\rm U}_d$ defines a twirl operation with the
property $\rho\in\wss\iff \pr\rho=\rho$.

Up to here all statements generalize easily to an arbitrary number
of factors, and some are even valid for an arbitrary averaging
operation with respect to a compact symmetry group. However, to
carry the analysis further one needs at least a precise
description of the range of the coefficients $\mu_\pi$ in equation
(\ref{sumpiV}), such that the sum indeed represents a density
operator. In general this is difficult, although sometimes
\cite{DCT99} one even gets a simplex. In the case we study in this
paper, the positivity of (\ref{sumpiV}) is best seen by using the
following basis, rather than the set of permutation operators
themselves.

\begin{mathletters}\label{V2R}
\begin{eqnarray}
  R_+\! &=& \frac{1}{6} \bigl(\idty+\V12 +\V23 +\V31 +\V123 +\V321 \bigr)\\
  R_-\! &=& \frac{1}{6} \bigl(\idty-\V12 -\V23 -\V31 +\V123 +\V321 \bigr)\\
  R_0 &=& \frac{1}{3} \Bigl(2\cdot\idty- \V123 -\V321 \Bigr)\\
  R_1 &=& \frac{1}{3} \Bigl(2\V23 -\V31 -\V12 \Bigr)\\
  R_2 &=& \frac{1}{\sqrt3}\ \Bigl(\V12 -\V31 \Bigr)\\
  R_3 &=& \frac{i}{\sqrt3}\ \Bigl(\V123 -\V321 \Bigr)\;,
\end{eqnarray}
\end{mathletters}
where we have used cycle notation to represent permutations. Then
$R_+,$ $R_-$ and $R_0$ are orthogonal projections commuting with
all $V_\pi$, and adding up to one. The $R_i$ for $i=1,2,3$ fulfill
the Pauli commutation relations $\ko{R_i}{R_\pm}=\Null,$
$R_i^2=R_0$, and $R_1R_2=iR_3$ and cyclic. Now every operator
$\rho$ in the linear span of the permutations can be decomposed
into the orthogonal parts $R_+\rho$, $R_-\rho$, and $R_0\rho$, and
positivity of $\rho$ is equivalent to the positivity of all three
operators. This leads to the following Criterion:

\begin{Cri}\label{le:range:ri}For any operator $\rho$ on
$\Hh\otimes\Hh\otimes\Hh$, define the six parameters
$r_k(\rho)=\tr(\rho R_k)$, for $k\in\lbrace+,-,0,1,2,3\rbrace$.
Then $r_k(\pr\rho)=r_k(\rho)$. Moreover, each $\rho\in\wss$ is
uniquely characterized by the tuple
$(r_+,r_-,r_0,r_1,r_2,r_3)\in\Rl^6$, and such a tuple belongs to a
density matrix $\rho\in\wss$ if and only if
\begin{eqnarray}
  r_+,r_-,r_0\geq0,&\qquad & r_++r_-+r_0=1 \nonumber\\
 {\rm and} &\qquad & r_1^2+r_2^2+r_3^2\leq r_0^2.\label{eq:range:ri}
\end{eqnarray}
\end{Cri}
Taking $r_0=1-r_+-r_-$ to be redundant, we get a simple
representation of $\wss$ as a convex set in $5$ dimensions.

Note that in this parametrization the set $\wss$ does not depend
on the dimension $d$ with one exception: for $d=2$ the
anti-symmetric projection $R_-$ is simply zero, so for qubits we
get the additional constraint $r_-=0.$ Although in the case of three
Qubits the dimensionality of the set is the same as in
\cite{DCT99} the two sets differ. In fact the set presented in
\cite{DCT99} can be obtained by averaging over a group of local unitaries of order 24.

With this parametrization
we can describe the 5 dimensional set $\wss$ by the points lying
in the $(r_+,r_-)$ plane together with a point $(r_1,r_2,r_3)$ in
the corresponding Bloch sphere of radius $r_0=1-r_+-r_-$. We note
that a state $\rho\in\wss$ is invariant under cyclic permutations
iff $r_1=r_2=0$, invariant under the interchange
$2\leftrightarrow3$ iff $r_2=r_3=0$, and invariant under all
permutations iff $r_1=r_2=r_3=0$. The latter set will be denoted
by $\wssp$.

\section{Results}

The basic results are summarized in Figure 1. On the one hand,
each point in this triangle corresponds to a density matrix
$\rho\in\wss$ with $r_1=r_2=r_3=0$, i.e., a permutation invariant state.
On the other hand, each such
point stands for the collection of states with the specified
$(r_+,r_-)$, and arbitrary $(r_1,r_2,r_3)$, all of which are
projected to the state with vanishing $(r_1,r_2,r_3)$ upon permutation
averaging. For qubits, one always has $r_-=0$, so only the
abscissa EB remains. Otherwise all statements are valid in any
dimension $d$.

\begin{figure}
\centerline{
\epsfxsize=\hsize
\epsffile{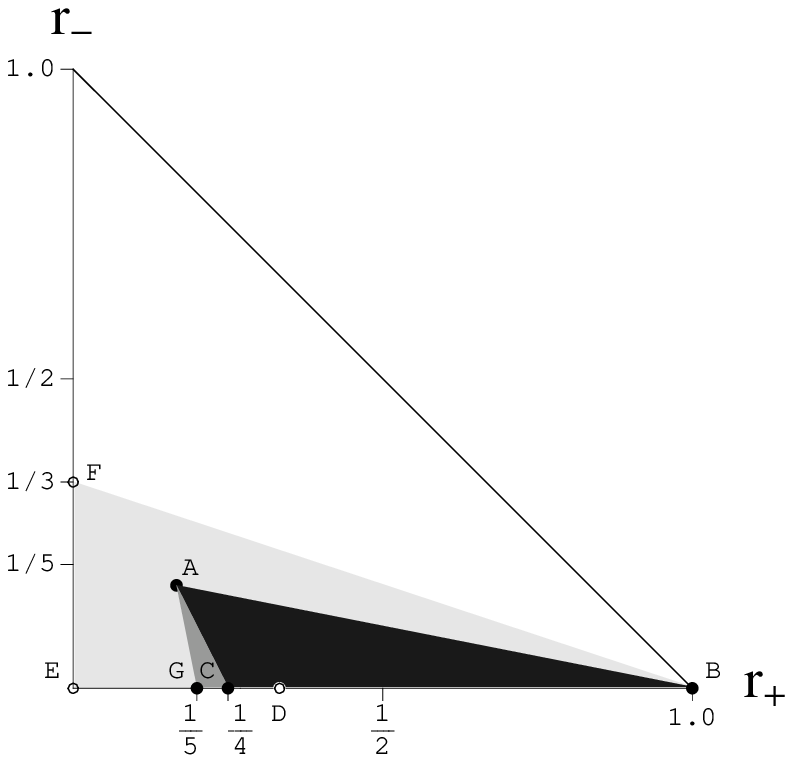}}
\caption{The permutation invariant subset $\wssp\subset\wss$.
Black: $\tsp\cap\wssp$; black or dark grey: $\bsp_1\cap\wssp$;
light grey: projection of $\bsp_1$ onto $\wssp$ by permutation
averaging. Marked points: see text.}\label{f1}
\end{figure}

\subsection{Triseparable States}
The black triangle ABC in Figure~1 is the set of
triseparable states. Its
vertices are obtained by taking suitable pure product states
$\ketbra{\Psi}{\Psi}$ with $\Psi=\phi_1\otimes\phi_2\otimes\phi_3$,
($\norm{\phi_i}=1$), and
applying the averaging projection $\pr$. Then one gets the points
A,B,C, if for $i\neq j$
\begin{eqnarray}
A&\mathpunct:&\quad \braket{\phi_i}{\phi_j}=0\\
B&\mathpunct:&\quad \braket{\phi_i}{\phi_j}=1\\
C&\mathpunct:&\quad \braket{\phi_i}{\phi_j}=\cos(2\pi/3)\;.
\end{eqnarray}
Note that the ``Mercedes Star'' configuration for $C$ requires
only two dimensions, whereas $A$ requires $d\geq3$. The technique
for getting all of $\tsp$ is similar: an arbitrary pure product
state is projected into $\wss$, and $\tsp$ is computed as the
convex hull of these points. This is more difficult than it
sounds, and the details will be presented elsewhere. The result
can be summarized as follows.

\begin{Cri}
A state $\rho\in\wss$ is triseparable if and only
if the following inequalities are satisfied:
\begin{itemize}
\item[\rm (a)] $0\leq r_- \leq \frac{1}{6}$
\item[\rm (b)] $\frac{1}{4}(1-2r_-)\leq r_+ \leq (1-5r_-)$
\item[\rm (c)] $(3r_3^2+[1-3 r_+]^2)\cdot(1-6r_-)\leq \\
\hspace*{1cm} (r_1+r_+-r_-)\cdot\left((r_1-2r_+ +2r_-)^2 -
3r_2^2\right).$
\end{itemize}
\end{Cri}

All extreme points other than $A=(\frac16,\frac16,0,0,0)$ are in
the plane $r_-=0$, i.e., they can also be realized by three
qubits. For fixed $(r_+,r_-)$, the shape of $\tsp$, as embedded in
the Bloch sphere parametrized by $(r_1,r_2,r_3)$ is shown in
Figure~2 (innermost convex set). Its threefold symmetry is the
residue of the permutations of the three sites.

\subsection{Biseparable States}
We fix the partition $1\vert 23$ of the system. Since the
projection of permutation averaging does not preserve
biseparability with respect to this partition, we now have to
distinguish two sets in Figure~1: those which are biseparable
states with $r_1=r_2=r_3=0$, hence permutation invariant
(represented as dark grey or black), and those which are the
images of some biseparable state under permutation averaging
(represented as light grey). So a light grey point in Figure~1 has
the property that for some suitable $(r_1,r_2,r_3)$ one gets a
biseparable state. Special points with this property are
represented by white circles, as opposed to filled circles, which
lie in the plane.

One interesting point in Figure~1 is $G=(\frac15,0)$. It is
biseparable and also permutation invariant. In particular, it is
biseparable for {\it any} partition of the system. But it is not
triseparable and, in fact, the only extreme point of the
permutation invariant biseparable set, which is not triseparable.
It can be obtained by applying $\pr$ to the pure state with vector
$\Psi=(\ket{112}-\ket{121}-\sqrt{3}\ket{122})/\sqrt{5}$.

The basic technique for computing $\bsp_1$ is the same as in the
triseparable case: one takes pure states with vectors of the form
$\Psi=\phi_1\otimes\phi_{23}$, and computes the convex hull of
their images under $\pr$. Some special extreme points of $\bsp_1$
are given below. They have the additional property of being
invariant under the exchange of systems $2$ and $3$, which is
equivalent to $r_2=r_3=0$. The following table lists the tuples
$(r_+,r_-,r_1)$, and the vector $\Psi$, in a suitable basis.
\begin{eqnarray}
B\mathpunct:&(1,0,0)&\qquad \ket{111}\\
D\mathpunct:&(\frac13,0,\frac23)&\qquad \ket{122}\\
E\mathpunct:&(0,0,-1)&\qquad (\ket{112}-\ket{121})/\sqrt{2}\\
F\mathpunct:&(0,\frac13,-\frac23)&\qquad (\ket{123}-\ket{132})/\sqrt{2}\;.
\end{eqnarray}

In addition to these four points there is a sphere of extreme
points extending also into the $r_2$ and $r_3$-dimensions, which
is tangent to the line connecting $D$ and $E$. The inequalities
describing the biseparable set $\bsp_1$ are given in the following

\newpage
\begin{Cri}
A state $\rho\in\wss$ is biseparable with respect to the partition $1\vert23$
if and only if $0\leq r_- \leq \frac{1}{3}$, and one of the
following conditions holds:
\begin{itemize}
\item[\rm (a)] $(3r_--1)\leq(1+r_1-r_--2r_+)\leq 0$ and
\begin{eqnarray*}
3r_2^2+3r_3^2+(1+2r_1^{}&+&r_-^{}-r_+^{})^2\\
&\leq& (2+r_1^{}-4r_-^{}\!-2r_+^{})^2.
\end{eqnarray*}
\item[\rm (b)] $0\leq(1+r_1-r_--2r_+)\leq(1-3r_-)$ and
$$3r_2^2+3r_3^2+(1-3r_-^{}-3r_+^{})^2\leq (r_1^{}+2r_-^{}-2r_+^{})^2.$$
\end{itemize}
\end{Cri}

For a typical point in Figure~1, the subset $\bsp_1$ in the Bloch
sphere is depicted in Figure~2. Note that the boundary is composed
of two quadratic surfaces, corresponding to the two alternatives
in the above criterion.

\begin{figure}
\centerline{
\epsfxsize=\hsize
\epsffile{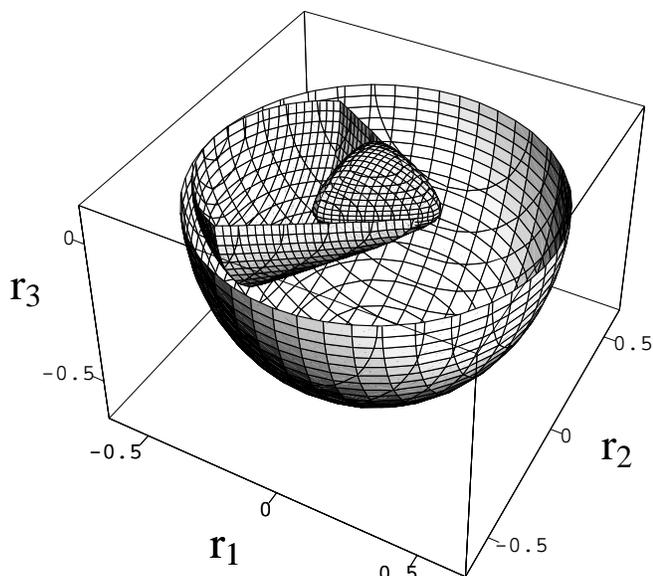}}
\caption{The Bloch sphere
over the point $r_+=0.27$ and $r_-=0.1$ in $\wssp$. Innermost
solid: triseparable states $\tsp$. Between $\tsp$ and outer
sphere: the boundary of the biseparable states
$\bsp_1$.}\label{f2}
\end{figure}

\subsection{States with Positive Partial Transpose}
Holding the partition $1\vert 23$ fixed we can compare the set of
states with positive partial transpose with respect to the first
subsystem ($\pps_1$) to $\bsp_1$. If Peres' Criterion were valid
in this case, we would have equality in $\ppt_1\supset\bsp_1$. It
turns out that the inclusion is strict, but in several respects
the criterion is amazingly good. To begin with, the intersections
of both sets with two important hyperplanes coincide: namely (1)
the $r_-=0$ plane (in particular, for three qubits),  and (2) the
$r_2=r_3=0$ plane, i.e. for states which are invariant under the
$2\leftrightarrow3$ interchange. In particular, the projections to
the permutation invariant subset coincide, so no difference can be
seen  in Figure~1.

Despite this similarity the technique for computing $\pps_1$ is
completely different. The key is the observation that partial
transposition in the first factor maps operators commuting with
all unitaries $(U\otimes U\otimes U)$ to operators commuting with
all unitaries of the form $(\overline{U}\otimes U\otimes U)$.
Obviously, the latter set is again an algebra, which even happens
to be isomorphic to the algebra $(U\otimes U\otimes U)$-invariant
operators: two one-dimensional summands plus the
$2\times2$-matrices. Hence positivity of partial transposes can be
decided along the same lines as in Criterion~1.

\begin{Cri}\label{Lem:ppt}Let $\rho\in\wss$ be a density operator
with expectations $r_k=\tr(\rho R_k)$, $k=+,-,1,2,3$. Then the
partial transpose of $\rho$ with respect to the first tensor
factor is positive, i.e. $\rho\in\pps_1$, if and only if
\begin{mathletters}\label{thm:ppt}
\begin{eqnarray}
s_1\equiv 1-r_1-5r_--r_+   &\geq&0\\
s_2\equiv -1-r_1+r_-+5 r_+ &\geq&0\\\text{\em and}\qquad
r_2^2+r_3^2&\leq&s_1s_2/3\;.\label{thm:ppt:e}
\end{eqnarray}
\end{mathletters}
\end{Cri}

Note that condition (\ref{thm:ppt:e}) is exactly the same as the
quadratic inequality in Criterion~3, Part(b). Therefore, we need
not even provide a new plot for the set $\ppt_1$: In Figure~2,
this set can be obtained simply by extending the quadratic
surface, which wraps around $\tsp$, all the way to the boundary of
the Bloch sphere. In other words, the difference between $\bsp_1$
and $\ppt_1$ is only that states in $\bsp_1$ have to satisfy an
additional quadratic inequality, which is represented in Figure~2
by the surface tangent to the Bloch sphere.

\section*{Acknowledgements}
We would like to thank M. Horodecki for discussions and the
Deutsche Forschungsgemeinschaft (DFG) for supporting this work.

\end{document}